\documentclass[final,5p,times,twocolumn, sort&compress]{elsarticle}

\usepackage{graphicx}
\usepackage{amssymb}
\usepackage{textcomp}
\usepackage{wasysym}
\usepackage{lineno}
\usepackage{epstopdf}
\usepackage{natbib}
\usepackage{hyperref}
\hypersetup{colorlinks = true}

\journal{Phys. Lett. B}

\begin{document}

\begin{frontmatter}

\title{The proton charge radius extracted from the Initial State Radiation experiment at MAMI}

\author[FMF,JSI,KPH]{M.~Mihovilovi\v{c}}
\author[KPH]{P.~Achenbach}
\author[KPH]{T.~Beranek}
\author[JSI]{J.~Beri\v{c}i\v{c}}
\author[MIT]{J.~C.~Bernauer}
\author[KPH]{R.~B\"{o}hm}
\author[Zagreb]{D.~Bosnar}
\author[KPH]{M.~Cardinali}
\author[IN2P3]{L.~Correa}
\author[JSI]{L.~Debenjak}
\author[KPH]{A.~Denig}
\author[KPH]{M.~O.~Distler}
\author[KPH]{A.~Esser}
\author[KPH]{M.~I.~Ferretti~Bondy}
\author[IN2P3]{H.~Fonvieille}
\author[TUM]{J.~M.~Friedrich}
\author[MIT]{I.~Fri\v{s}\v{c}i\'{c}}
\author[CWM]{K.~Griffioen}
\author[KPH]{M.~Hoek}
\author[KPH]{S.~Kegel}
\author[KPH]{Y.~Kohl}
\author[KPH]{H.~Merkel\corref{cor1}}
  \ead{merkel@kph.uni-mainz.de}
\author[KPH]{D.~G.~Middleton}
\author[KPH]{U.~M\"{u}ller}
%\author[KPH]{L.~Nungesser}
\author[KPH]{J.~Pochodzalla}
%\author[KPH]{M.~Rohrbeck}
%\author[KPH]{S.~S\'anchez~Majos}
\author[KPH]{B.~S.~Schlimme}
\author[KPH]{M.~Schoth}
\author[KPH]{F.~Schulz}
\author[KPH]{C.~Sfienti}
\author[FMF,JSI]{S.~\v{S}irca}
\author[JSI]{S.~\v{S}tajner}
\author[KPH]{M.~Thiel}
\author[KPH]{A.~Tyukin}
\author[KPH]{M.~Vanderhaeghen}
\author[KPH]{A.~B.~Weber}
%\author[KPH]{M.~Weinriefer}

\cortext[cor1]{Corresponding author}

\address[FMF]{Faculty~of~Mathematics~and~Physics, University~of~Ljubljana, SI-1000 Ljubljana, Slovenia}
\address[JSI]{Jo\v{z}ef~Stefan~Institute, SI-1000 Ljubljana, Slovenia}
\address[KPH]{Institut~f\"{u}r~Kernphysik, Johannes~Gutenberg-Universit\"{a}t~Mainz, DE-55128~Mainz,~Germany}
\address[MIT]{Massachusetts Institute of Technology, Cambridge, MA~02139, USA}
\address[Zagreb]{Department~of~Physics, Faculty~of~Science, University~of~Zagreb, HR-10002~Zagreb, Croatia}
\address[IN2P3]{Universit\'{e}~Clermont~Auvergne, CNRS/IN2P3, LPC, BP~10448, F-63000 Clermont-Ferrand, France}
\address[TUM]{Technische Universit\"{a}t M\"{u}nchen, Physik Department, 85748 Garching, Germany}
\address[CWM]{College of William and Mary, Williamsburg, VA~23187, USA}

\begin{abstract}
We report on a comprehensive reinterpretation of the existing cross-section
data for elastic electron-proton scattering obtained by the initial-state
radiation technique, resulting in a significantly improved accuracy
of the extracted proton charge radius.  By refining the external energy
corrections we have achieved an outstanding description of the radiative tail,
essential for a detailed investigation of the proton finite-size effects
on the measured cross-sections. This development, together with a novel
framework for determining the radius, based on a regression analysis
of the cross-sections employing a polynomial model for the form factor,  
led us to a new value for the charge radius, which is
$(0.870 \pm 0.014_\mathrm{stat.}\pm 0.024_\mathrm{sys.} \pm
0.003_\mathrm{mod.})\,\mathrm{fm}$.

\end{abstract}

\begin{keyword}
Initial state radiation \sep Proton  radius \sep Radiative corrections
\PACS12.20.-m \sep 25.30.Bf \sep 41.60.-m
\end{keyword}

\end{frontmatter}



\section{Introduction}
The problem of the proton charge radius persists. The 
difference between the CODATA~\cite{CODATA2014} value of 
$0.8751(61)\,\mathrm{fm}$, compiled from electron scattering, and the
old atomic Lamb shift measurements differ significantly from the very precise 
Lamb shift measurements in muonic hydrogen~\cite{Pohl2010, Antognini2013}, 
which give a value of $0.84087(39)\,\mathrm{fm}$. 
Even with new measurements, the discrepancy remains unresolved. 
Although the measurement of the $2S$-$4P$ transition in Hydrogen~\cite{Beyer79} yields a 
value of $0.8335(95)\,\mathrm{fm}$, which is in concordance with the smaller radius,
the measurement of the $1S$-$3S$ transition~\cite{Fleurbaey2018} gives a value 
of $0.877(13)\,\mathrm{fm}$, consistent with the CODATA value. Therefore, additional
experiments, both scattering and spectroscopic,  have the potential to make valuable 
contributions to the proton size problem~\cite{PohlAnn,Carlson201559}.

In scattering experiments the charge radius of the proton is traditionally determined by 
measuring the cross section for elastic scattering of electrons from hydrogen, which 
depends on $G_E^p$ and carries information about the charge 
distribution of the proton. The proton charge radius, $r_p$,  is given by
\begin{eqnarray}
r_p^2 \equiv \left.-6\hbar^2 \frac{\mathrm{d}G_E^p}{\mathrm{d}Q^2}\right |_{Q^2=0}\,,\label{pr1}
\end{eqnarray}
where $Q^2$ is the negative square of the four-momentum transfer to the proton. 
The accuracy of the radius obtained in this manner is limited by the extent of existing data sets
($Q^2 > 0.004\,\mathrm{GeV}^2/c^2$), which governs the  level of  extrapolation 
of  $G_E^p$ needed to determine the slope at $Q^2=0$.  Hence, to further improve the existing 
results, measurements of $G_E^p$ need to be extended into the previously unmeasured 
region of  $Q^2\lesssim 0.004\,\mathrm{GeV}^2/c^2$.

Efforts to perform such measurements with the standard approaches are limited by the minimum
$Q^2$ accessible with the experimental apparatus at hand, predominantly due to the 
restrictions in the available electron beam energy and the minimum scattering angle. 
Therefore, a new experimental approach based on 
 initial state radiation has been introduced to extend the 
currently accessible $Q^2$ range, and allow for cross section measurements below 
$0.004\,\mathrm{GeV^2}/c^2$ with sub-percent precision by using information 
about the charge form factor that is implicit in the radiative tail of the elastic peak. 

\section{Initial-state radiation experiment}

\begin{figure}[ht]
\begin{center}
 \includegraphics[width=0.48\textwidth]{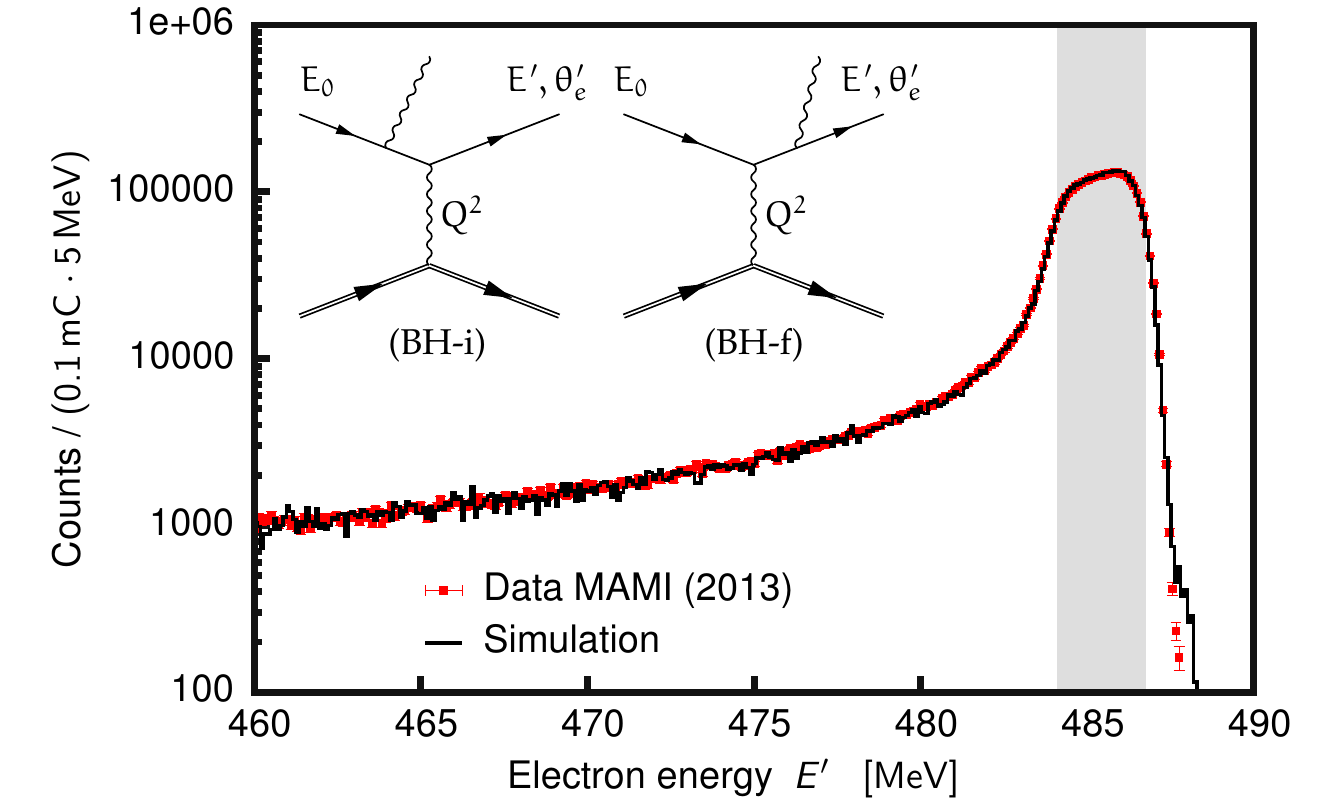}
 \caption{Measured and simulated elastic peak with the corresponding radiative tail 
 for the first kinematic setting at $495\,\mathrm{MeV}$. See~\cite{mihovilovic_PLB} for details. 
 The radiative tail is dominated by the two Bethe-Heitler diagrams (BH-i and BH-f), where 
 electrons emit real photons before or after the interaction with the protons. The grey band marks 
 the position and width of the elastic line inside the spectrometer acceptance.
 \label{fig_ISRDiagrams}}
 \end{center}
\end{figure}

The radiative tail of an elastic peak is dominated by the coherent sum of two Bethe-Heitler
diagrams~\cite{Vanderhaeghen2000} shown in Figure~\ref{fig_ISRDiagrams}. The initial state radiation 
diagram (BH-i) describes the process where the incident electron emits a real photon 
before interacting with the proton. Since the emitted photon carries away part of the incident 
energy, the momentum transfer to the proton  is decreased.
Hence, this process probes the proton structure at values of $Q^2$ smaller than the 
value fixed by the experimental kinematics and is thus sensitive to the form-factors at $Q^2$ smaller 
than those corresponding to the elastic setting. On the other hand, the final state radiation diagram (BH-f) 
corresponds to the reaction where the real photon is emitted after the interaction with the 
nucleon. Consequently, $Q^2$ at the vertex remains constant, while the detected four-momentum 
transfer changes.

In an inclusive experiment $Q^2$ can not be measured directly, which means 
that looking only at data the initial state radiation processes can not be distinguished from 
the final state radiation. Hence, in order to get information on $G_E^{p}$ at $Q^2$
smaller than  the elastic setting, the data must be studied in conjunction with a Monte-Carlo 
simulation, which includes a detailed description of the radiative corrections and considers $G_E^{p}$ as its
free parameter. This is the basic idea of the MAMI experiment, which opened the door 
of obtaining $G_{E}^{p}$ down to 
$Q^2 \simeq 10^{-4}\,\mathrm{GeV^2}/c^2 $~\cite{mihovilovic_PLB}.

The measurement of 
the radiative tail has been performed at the Mainz Microtron (MAMI) in 2013  using 
the spectrometer setup of the A1-Collaboration~\cite{Blomqvist}. 
A rastered electron beam with energies of $E_0=195$, 
$330$ and $495\,\mathrm{MeV}$ was used in combination with a 
hydrogen target, which consisted of a $5\,\mathrm{cm}$-long cigar-shaped 
Havar cell filled with liquid hydrogen and
placed in an evacuated scattering chamber. For the cross section measurements 
the single-dipole magnetic spectrometer~B was employed at a fixed 
angle of $15.21^\circ$, while its momentum settings were adjusted to scan 
the complete radiative tail for each beam energy. The central momentum of each 
setting was measured with an NMR probe to a relative accuracy of $8\times10^{-5}$. 
The spectrometer was equipped with the standard detector package consisting of two layers of 
vertical drift chambers (VDCs) for tracking, two layers of scintillation detectors 
for triggering, and a threshold Cherenkov detector for particle identification. 
The kinematic settings of the experiment were chosen such that the radiative 
tails scanned at three beam energies overlap. 

The beam current was between $10\,\mathrm{nA}$ and $1\,\mu\mathrm{A}$
and was limited by the maximum rate allowed  in the VDCs 
($\approx 1\,\mathrm{kHz/wire}$), resulting in raw rates 
up to $20\,\mathrm{kHz}$. The beam current was determined by 
a non-invasive fluxgate-magnetometer and from the collected charge of the beam
stopped in a Faraday cup. At low beam currents and low beam energies the accuracy of
both approaches is not better than $ 2\,\mathrm{\%}$, which is insufficient for 
precise cross section measurements. Hence Spectrometer~A was used at a fixed 
momentum and angular setting for precise monitoring of the 
relative luminosity.

\begin{figure}[ht]
\begin{center}
 \includegraphics[width=0.50\textwidth]{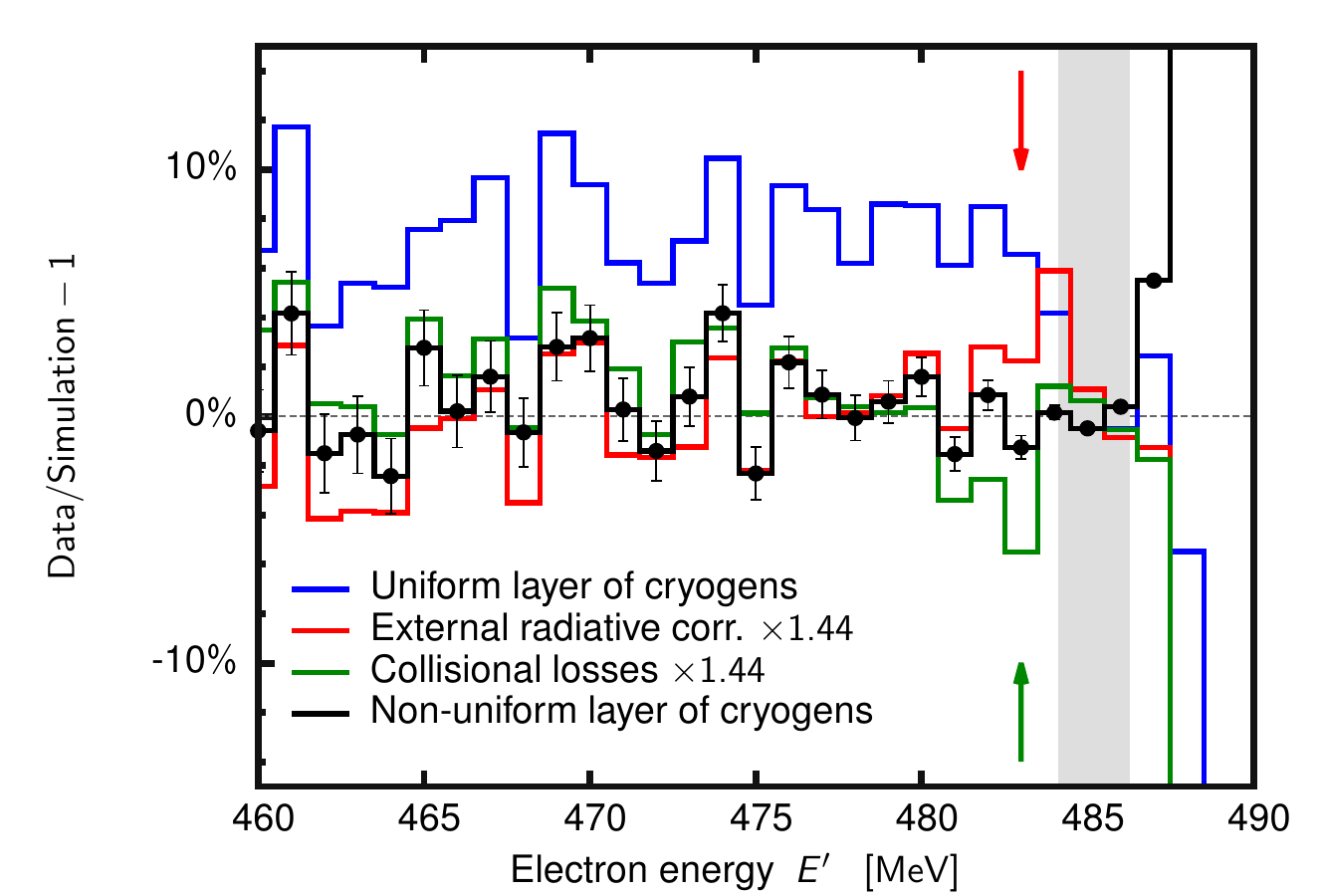}
 \caption{
 Relative differences between the data and simulations 
 for the first experimental setup at $495\,\mathrm{MeV}$, including
 data at the elastic peak and first part of the radiative tail. 
 The blue line shows the original comparison considered in ~\cite{mihovilovic_PLB},
 when the simulation assumes a uniform layer of cryogens around the target 
 cell. The red line corresponds to the result with the artificially amplified 
 external radiative correction given by Mo-Tsai~\cite{MoTsai}. The green line 
 shows the result achieved with the simulation with enhanced 
 collisional corrections described by the Landau distribution. The black line shows 
 the result with the simulation using an unmodified description of the external corrections 
 but considering that away from the beam, the layer of cryogens is approximately 
 200 times thicker than along the beam. The red (green) 
 arrow indicates the position of the surplus (deficiency) of events, when using
 modified radiative (collisional) losses. The systematic 
 uncertainty of all the points is $0.4\,\mathrm{\%}$. The grey band marks 
 the position and width of the elastic line inside the spectrometer acceptance. \label{fig_Snow}}
 \end{center}
\end{figure}

The analysis of the data, presented in \cite{mihovilovic_PLB}, revealed 
inconsistencies between data and simulation on the 
order of $10\,\mathrm{\%}$ at the top of the
elastic peak, see Fig.~\ref{fig_Snow}, which led to the omission 
of the most statistically relevant,  elastic data points in the sample.  
We believed that the inconsistency arose due to the incomplete theoretical description 
of the external radiative corrections in the target material in the limit of very small 
photon energies, which were not considered to the same  order of precision as the 
internal corrections. To be able to incorporate the elastic data in the analysis and ensure 
a more precise extraction of the proton charge radius, the sensitivity of the results to the 
applied external radiative corrections and collisional energy losses has been investigated 
in detail and is shown in Fig.~\ref{fig_Snow}. 

First, the external radiative correction, given by Mo and Tsai~\cite{MoTsai},
were artificially amplified by a factor of $1.44$. This modification brought the simulation into agreement
with the data at the elastic peak and below $E'=480\,\mathrm{MeV}$, but caused a reduction of 
simulated events in the first $5\,\mathrm{MeV}$ of the  radiative tail. A similar result was observed with 
the artificial enhancement of collisional corrections described by the Landau distribution, 
but this time with an excess of simulated events just below the elastic peak. Hence, if both 
corrections are changed simultaneously, %by the same amount,  
a consistent description of the radiative 
tail and a correspondingly small relative difference over the whole momentum range
are observed. This finding suggests that the observed inconsistency is a consequence 
of an unaccounted for material traversed by the scattered electrons  and is not due to an incomplete 
description of the applied external corrections. This was confirmed by 
the dedicated followup experiment, using the same experimental setup but 
different targets. Data were collected using plastic ($[\mathrm{CH_2}]_n$) 
targets with different thicknesses, which created a perfect testbed for
validating the applied corrections. The analysis of the elastic peak shapes 
demonstrated a good agreement between the data and simulations and exhibited the correct 
scaling of the corrections with the thickness of the target. 

In spite of the good vacuum conditions inside the scattering chamber 
($10^{-6}\,\mathrm{mbar}$), the experiment was sensitive to traces of 
cryogenic deposits on the target walls, consisting mostly of residual 
nitrogen and oxygen present in the scattering chamber~\cite{MihovilovicMessina}. 
To determine the amount of cryogenic deposits, spectrometer A was 
positioned such that the nitrogen/oxygen elastic lines were inside its acceptance. 
The collected spectra, together with the known elastic cross-sections for these
elements were used to determine the thickness of the deposited layer. However, 
the analysis assumed that the layer has the same thickness in all parts of the 
target cell, disregarding that the depositions on the side walls can be much thicker 
than those on the end caps, because the former are not heated by the electron beam.  
Since the material on the side walls is not detectable by the spectrometers, the amount of 
cryogens was estimated by matching the functional dependence of the simulation to 
the measured elastic spectra. See Fig.~\ref{fig_Snow}. This showed that the layer 
of cryogens on the side can be as much as $200$ times thicker (roughly $4\cdot10^{-3}\,\mathrm{g/cm^2}$) 
than at the end caps. This new insight significantly improved the agreement 
between the data and simulation and allowed us to include the elastic data in a new analysis.

\section{Experimental uncertainties}
Although the ISR experiment provides remarkable control over the
systematic uncertainties, a few ambiguities remain and limit the precision 
of the results~\cite{mihovilovic_PLB}.  The contributions relevant 
for the extraction of the proton charge radius include the uncertainty in the 
relative luminosity ($0.17\,\mathrm{\%}$), the ambiguity 
in the detector efficiencies $(0.2\,\mathrm{\%})$ and
the contamination  coming from the target support frame and 
the spectrometer entrance flange $(0.4\,\mathrm{\%})$. 
The portion of the spectrum containing contributions from the pion 
electroproduction is $0.5\,\mathrm{\%}$, but is 
significant only for the $495\,\mathrm{MeV}$ setting. The combined point-wise systematic 
uncertainties are presented in Fig.~\ref{fig_Ratios}.

\section{Extraction of the radius}

Following the approach described in~\cite{mihovilovic_PLB} the full set of 
$25$ data points was used to extract the proton-charge form factors for  
 $0.001 \leq  Q^2 \leq 0.017\,\mathrm{GeV}^2/c^2$. The values and the details 
 of the extraction are presented in~\cite{supply}. These were
 then compared to the polynomial function  
  \begin{equation}
G(Q^2)= n_{E_0}\left[1 - \frac{r_p^2\,Q^2}{6\,\hbar^2}  + \frac{a\,Q^4}{120\,\hbar^4} - 
   \frac{b\,Q^6}{5040\,\hbar^6}\right] \,, \label{eq2}
\end{equation}
 where parameters $a=(2.59\pm0.194)\,\mathrm{fm}^4$ and $b=(29.8\pm14.71)\,\mathrm{fm}^6$, 
which determine the curvature of the fit, were taken from Ref.~\cite{distler}.
The three data sets were fit with a common parameter for the radius, $r_p$, 
but with different renormalisation factors, $n_{E_0}$, for each energy.
In terms of this fit with $21$ degrees of freedom and $\chi^2$ of $95.4$, 
the normalisations and the radius were determined to be:
\begin{eqnarray} 
n_{195} &=&1.002 \pm 0.002_{\mathrm{stat}} \pm 0.007_{\mathrm{syst}}\,, \nonumber\\
n_{330} &=&1.000 \pm 0.001_{\mathrm{stat}} \pm 0.003_{\mathrm{syst}}\,, \nonumber\\
n_{495} &=&0.999 \pm 0.001_{\mathrm{stat}} \pm 0.004_{\mathrm{syst}}\,, \nonumber\\
r_p &=& (0.836 \pm 0.017_{\mathrm{stat}} \pm 0.059_{\mathrm{syst}} \pm 0.003_{mod})\,\mathrm{fm}\,. \nonumber
\end{eqnarray} 
The value of $\chi^2$ has almost doubled with respect to the previous dataset~\cite{mihovilovic_PLB}, 
but this is due to the addition of three statistically very precise points at $Q^2 = 0.017\,\mathrm{GeV^2}/c^2$,
$0.008\,\mathrm{GeV^2}/c^2$ and $0.003\,\mathrm{GeV^2}/c^2$. On the other
hand, even by including these points and improving the analysis in the above manner,
the extracted radius remains governed by the systematic uncertainty and still critically 
depends on the available $Q^2$ range and the number of fitting parameters. 

\begin{figure}[h]
\begin{center}
 \includegraphics[width=0.45\textwidth]{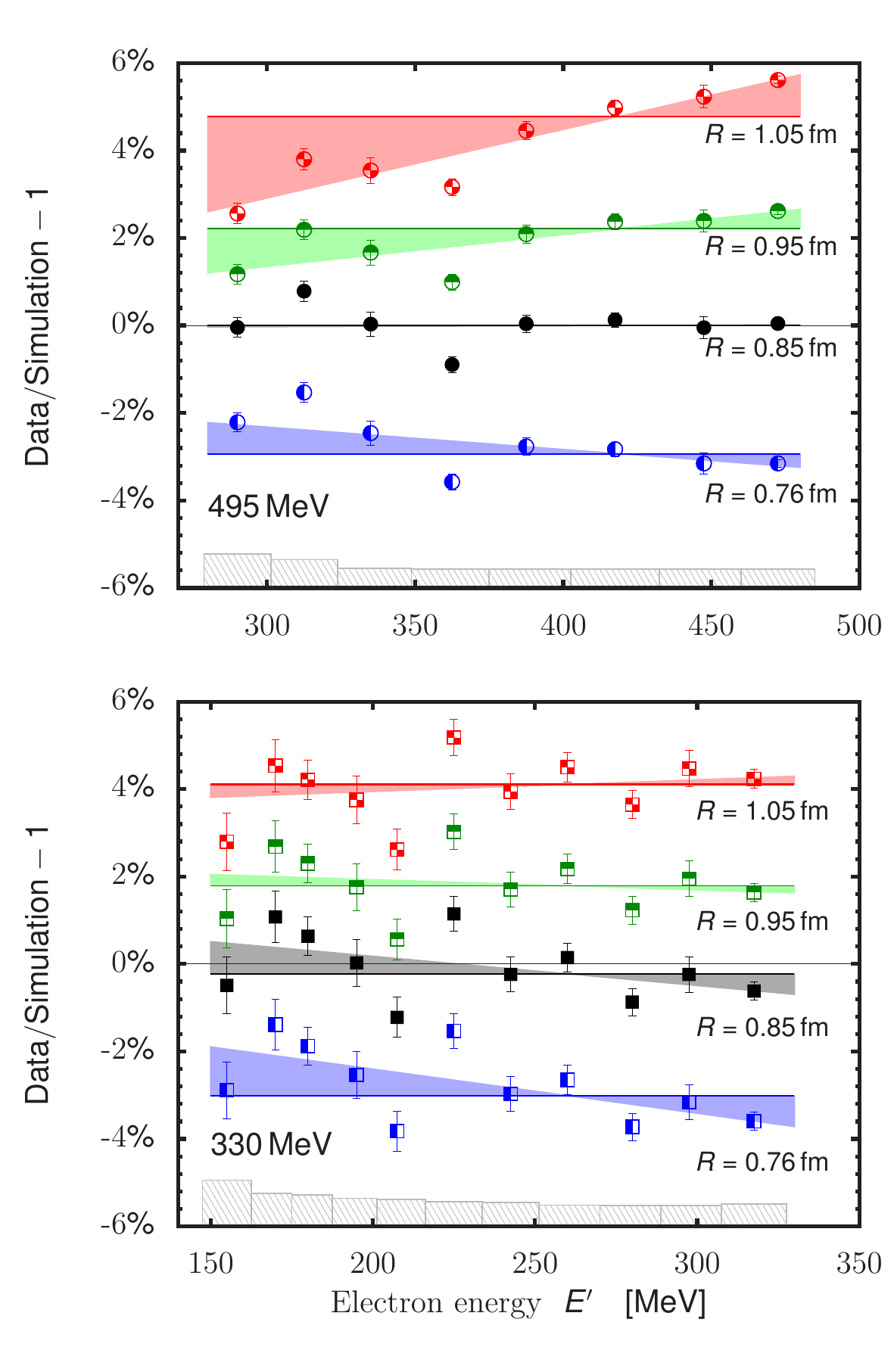}
 \caption{
  Relative differences between the data and simulations $R_{i(j)} (r_p)$ for $495\,\mathrm{MeV}$ (top)
  and $330\,\mathrm{MeV}$ (bottom) settings. Each set of ratios
  corresponds to simulations with a different value of the proton charge radius 
  $r_p = 0.76\,\mathrm{fm}$ (blue), $0.85\,\mathrm{fm}$ (black), $0.95\,\mathrm{fm}$ (green) 
  and $1.05\,\mathrm{fm}$ (red). The weighted averages $\overline{R}_{330}(r_p)$ and 
  $\overline{R}_{495}(r_p)$ are  shown with full lines and are offset from zero for clarity. The sizes of the
  corresponding triangular shapes denote the magnitude of the inconsistency between the 
  data and simulation.  Gray boxes demonstrate the systematic uncertainties. 
 \label{fig_Ratios}}
 \end{center}
\end{figure}

To improve these results within the scope of available data, an alternative approach
was considered, applicable at the level of measured cross-sections. 
To first order the $E'$ evolution of the ratio between the data and the simulation, $R_i(r_p)$,  
at each energy setting depends only on the proton charge radius. Here, index $i$ denotes 
the  kinematic setting with energy $E'$.  Furthermore, since all points for a single energy configuration are strongly correlated
 due to the nature of the experimental approach the 
effect of changing the radius appears as a change of the slope of the ratio. 
Hence, focusing on the slopes of the cross-section 
ratios effectively eliminates  normalisations from the minimisation procedure
and offers more accurate determination of the radius.    
To evaluate the sensitivity of the data to the proton charge radius, 
the simulation with the same form-factor model (\ref{eq2})  was 
run for 10 different values of $r_p$ between $0.76\,\mathrm{fm}$
and $1.05\,\mathrm{fm}$. 
The results of the comparison have demonstrated that 
only at the highest two energy settings ($330\,\mathrm{MeV}$ and 
$495\,\mathrm{MeV}$) the $Q^2$ is large enough for the measurements to be 
sensitive to small changes in the proton radius. See Fig.~\ref{fig_Ratios}.  
The data at $195\,\mathrm{MeV}$  exhibit no detectable dependence on the 
proton charge radius and, within the measured uncertainties, could be 
used only for the absolute calibration of the measured cross-sections.

\begin{figure}[ht]
\begin{center}
 \includegraphics[width=0.45\textwidth]{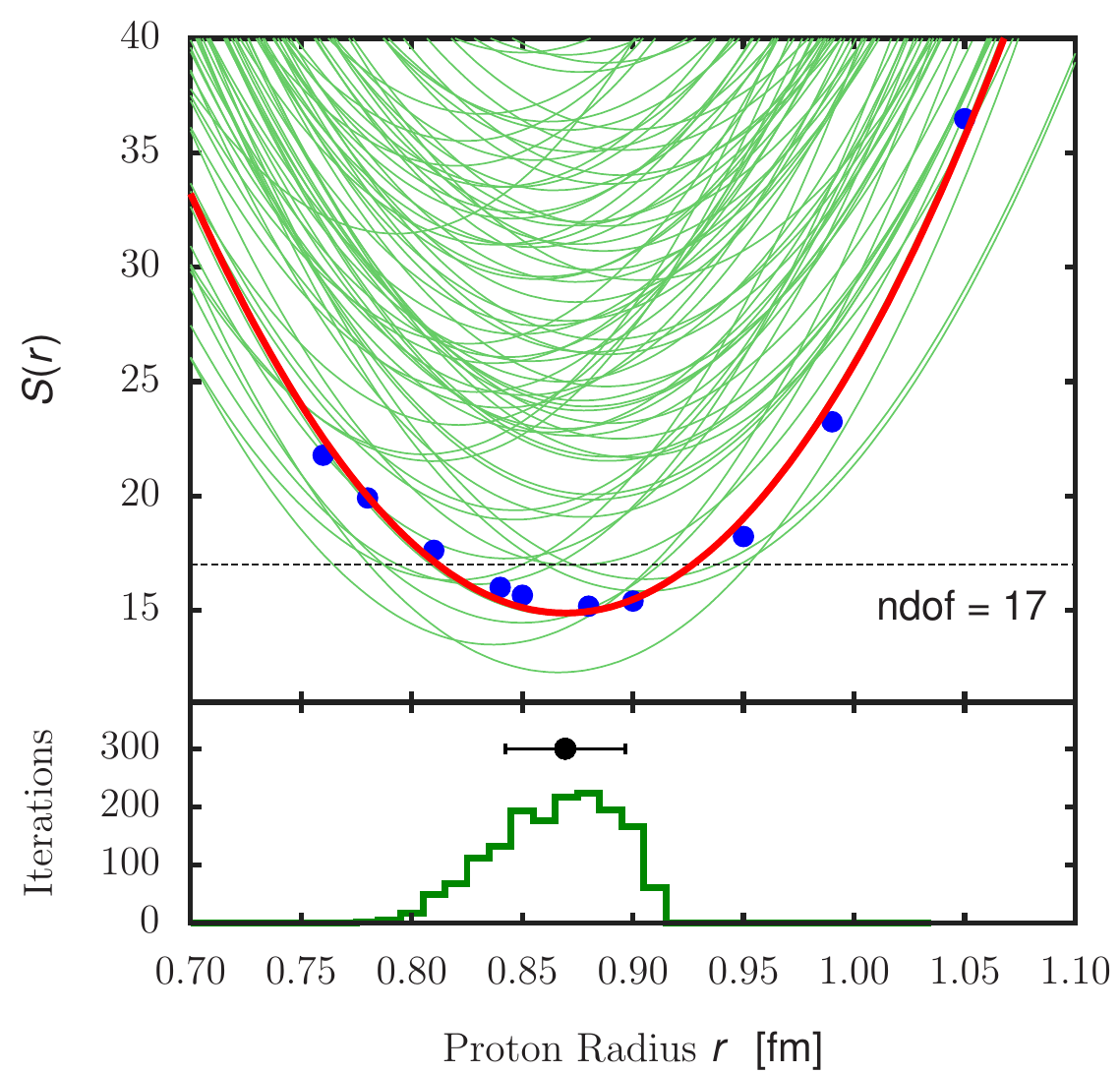}
 \caption{
Top: The blue circles show $S(r_p)$ corresponding to the comparison of the data to
 the simulations ran with $r_p = 0.76$, $0.78$, $0.80$, $0.84$, $0.85$, $0.88$, 
 $0.90$, $0.95$, $0.99$, and $1.05\,\mathrm{fm}$.  The red curve represents the parabola 
 fitted to the blue points. The horizontal line marks the number of degrees of freedom and 
 should ideally agree with the minimal value of the fit function. The green curves 
 represent parabolas, when data points are randomly smeared by their uncertainties. 
 Bottom: The histogram shows the distribution of minima of the parabolas shown in 
 the upper panel. The extracted radius presented with the black point corresponds 
 to the minimum of the red parabola in the upper panel. Its uncertainty is  determined by 
 the width of the distribution of minima.
 \label{fig_Minimization}}
 \end{center}
\end{figure}

Relying on the chosen model (\ref{eq2}), the best estimate for the proton charge radius 
should return  a flat ratio between the data and the simulation as a function of $E'$. 
The ratios obtained with 
different values of the parameter, presented in Fig.~\ref{fig_Ratios}, demonstrate 
that the $495\,\mathrm{MeV}$ setting favours a radius of $\approx 0.85\,\mathrm{fm}$, while 
the $330\,\mathrm{MeV}$ data suggest $\approx 0.95\,\mathrm{fm}$. To quantify the 
level of agreement between the data and the simulation performed with a particular value 
of $r_p$, the following function was considered:
\begin{eqnarray}
S(r_p) = \sum_{i=1}^{N_{495}}\frac{\left(R_i(r_p)-\overline{R}_{495}(r_p)\right)^2}{{\sigma_R}_i^2} +
\sum_{j=1}^{N_{330}}\frac{\left(R_j(r_p)-\overline{R}_{330}(r_p)\right)^2}{{\sigma_R}_j^2} \,. \nonumber
\end{eqnarray}
For a given value of $r_p$ the $\overline{R}_{330}(r_p)$ and $\overline{R}_{495}(r_p)$ 
represent weighted averages of ratios $R_{i}$ and $R_{j}$, while $N_{330}$ and $N_{495}$ correspond
to the number of measured points for $330\,\mathrm{MeV}$ and $495\,\mathrm{MeV}$ configuration, 
respectively.  The ${\sigma_R}_{i (j)}$ represent the total uncertainties of the
measured ratios. The values of $S(r_p)$ obtained for the selected radii are shown in 
Fig.~\ref{fig_Minimization}. Finding a radius that best matches both data sets amounts to 
finding the minimum of the function $S(r_p)$ . Unfortunately, the full numerical 
minimisation of the function could not be performed, because it would require
numerous repetitions of the full simulation, which is  computationally very intensive. 
Instead,  the simulated points were fitted by a parabola and the radius was determined
by finding its vertex at $r_p = 0.870\,\mathrm{fm}$ and 
$S = 14.89$.

Furthermore, $S(r_p)$ can be recognized as a $\chi^2$ estimator of the quality 
of the fit. One expects $N_{330}+N_{495}-2=17$ number of degrees of freedom, 
indicating that the fit reliably describes the data.

The uncertainty of the extracted radius was determined by means of a Monte-Carlo
simulation. In each iteration the extracted ratios $R_{i(j)}$ were changed randomly following 
a Gaussian distribution with a width given by ${\sigma_R}_{i(j)}$.  Additionally, since only one 
set of data is available and only the simulation is changing in the study, the points at the same 
energy (or index $i,j$) simulated with different radii are correlated. Hence, they
need to be offset simultaneously by the same amount. The applied changes result in a different 
minimisation parabola and different radius. Repeating the simulation many  
times, considering both statistical and systematic uncertainties, results in a distribution of 
radii, see Fig.~\ref{fig_Minimization} (bottom), whose width serves as an estimation of the 
uncertainty of the extracted proton charge radius.  The best value for 
the proton charge radius becomes:
$$
r_p = \left (0.870 \pm 0.014_\mathrm{stat.}\pm 0.024_\mathrm{sys.}\pm 0.003_\mathrm{mod.}\right)\,\mathrm{fm}\,.
$$  
The presented result contains a small model dependence on the function (\ref{eq2}) used
to describe $G_E^p$.  
The model uncertainty  arises from the uncertainties of the parameters $a$ and $b$,
which are robust and based on the data of Bernauer et al.~\cite{Berni2014}.  While 
our data are not precise enough for independent extraction of  $a$ and $b$,
the radius extraction procedure is sensitive to their specific value. Replacing $a$ and $b$ 
by those obtained from the ChPT calculations of Alarcon and Weiss~\cite{alarcon}, for instance 
would increase the function $S$, while the resulting radius would shift by about $-0.015\,\mathrm{fm}$ towards 
a smaller value. 

\section{Conclusions}

\begin{figure}[ht]
\begin{center}
 \includegraphics[width=0.46\textwidth]{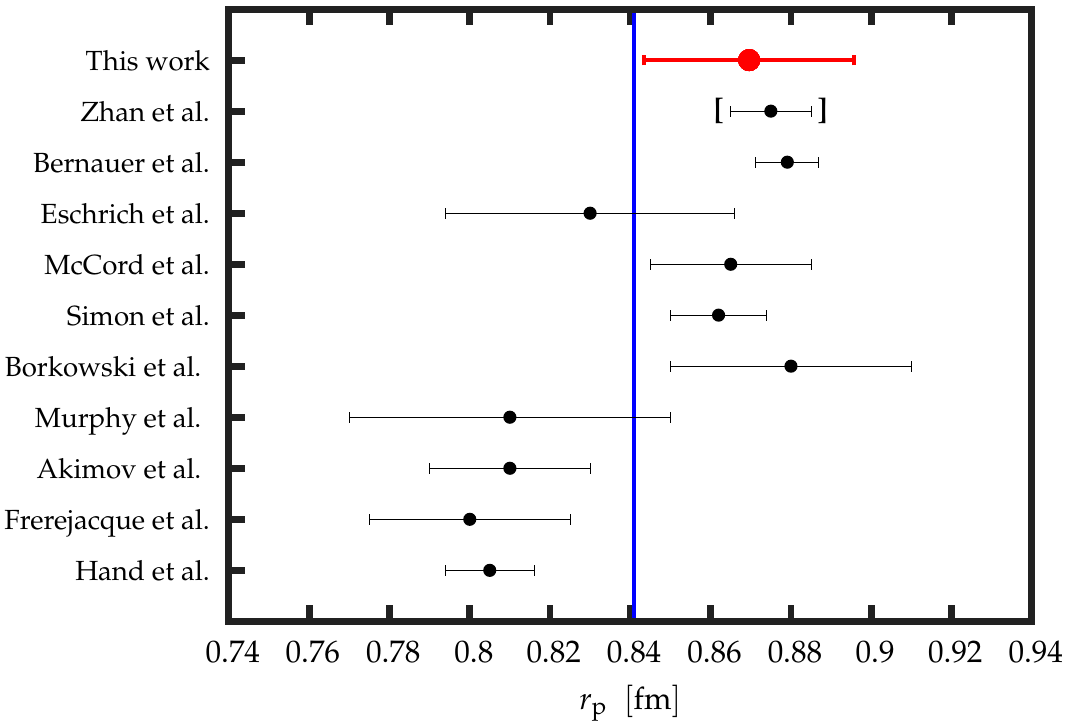}
 \caption{ The proton charge radius extracted from the ISR experiment together with 
previous electron scattering measurements~\cite{Hand, Frerejacque, Akimov, Murphy, Borkowski, SimonMAINZ, McCord, Eschrich, Bernauer2010}, where the data point of Zhan~et~al., enclosed by square 
brackets, is not an independent measurement of the radius~\cite{Zhan2011}. The value obtained from the  
 Lamb shift measurements in muonic hydrogen is shown by the blue line for the comparison. 
 \label{fig_Radius}}
 \end{center}
\end{figure}

The initial state radiation experiment at MAMI~\cite{mihovilovic_PLB} established a new method
for precise investigations of the electromagnetic structure of the nucleon and underlying 
electromagnetic processes at extremely small $Q^2$. In this paper we present our findings on the 
improved data analysis, which revealed the necessity of a  complete consideration 
of cryogens deposited on the liquid hydrogen cell and their influence on the $e$-$p$ scattering results. 
The analysis also demonstrated the precision with which these effects could be 
studied and offered new, improved values of the $G_E^p$ not accessible in the original work.  
Furthermore, by studying the slopes of the measured radiative tails relative to the simulated 
ones, an alternative approach for the extraction of the proton charge radius was developed, 
which yielded a competitive new value (see Fig.~\ref{fig_Radius}) 
and improved the precision of our initial result~\cite{mihovilovic_PLB} by almost a factor of $3$.

\section*{Acknowledgments} \label{acknowledgments}
The authors would like to thank the MAMI accelerator group for the excellent beam
quality which made this experiment possible. This work is supported by the 
Federal State of Rhineland-Palatinate, by the Deutsche Forschungsgemeinschaft 
with the Collaborative Research Center 1044, by the Slovenian Research 
Agency under Grant Z1-7305, by Croatian Science Foundation under the project 
IP-2018-01-8570 and U.~S. Department of Energy under Award Numbers 
DE-FG02-96ER41003 and DE-FG02-94ER40818.

%\section*{References}

\bibliographystyle{elsarticle-num} 
\bibliography{ISRPaperNo2PLB}

\end{document}